\documentclass[12pt]{article}
\usepackage{amsfonts}
\usepackage{latexsym}
\usepackage{times}

\catcode`\@=11

\newcount\hour
\newcount\minute
\newtoks\amorpm \hour=\time\divide\hour by 60\minute
=\time{\multiply\hour by 60 \global\advance\minute by-\hour}
\edef\standardtime{{\ifnum\hour<12 \global\amorpm={am}%
        \else\global\amorpm={pm}\advance\hour by-12 \fi
        \ifnum\hour=0 \hour=12 \fi
        \number\hour:\ifnum\minute<10
        0\fi\number\minute\the\amorpm}}
\edef\militarytime{\number\hour:\ifnum\minute<10
0\fi\number\minute}

\def\draftlabel#1{{\@bsphack\if@filesw {\let\thepage\relax
   \xdef\@gtempa{\write\@auxout{\string
      \newlabel{#1}{{\@currentlabel}{\thepage}}}}}\@gtempa
   \if@nobreak \ifvmode\nobreak\fi\fi\fi\@esphack}
        \gdef\@eqnlabel{#1}}
\def\@eqnlabel{}
\def\@vacuum{}
\def\marginnote#1{}
\def\draftmarginnote#1{\marginpar{\raggedright\scriptsize\tt#1}}
\def\draft{
        \pagestyle{plain}
        \overfullrule=2pt
        \oddsidemargin -.5truein
        \def\@oddhead{\sl \phantom{\today\quad\militarytime} \hfil
        \smash{\Large\sl DRAFT} \hfil \today\quad\militarytime}
        \let\@evenhead\@oddhead
        \let\label=\draftlabel
        \let\marginnote=\draftmarginnote
        \def\ps@empty{\let\@mkboth\@gobbletwo
        \def\@oddfoot{\hfil \smash{\Large\sl DRAFT} \hfil}
        \let\@evenfoot\@oddhead}
        \def\@eqnnum{(\theequation)\rlap{\kern\marginparsep\tt\@eqnlabel}%
        \global\let\@eqnlabel\@vacuum}  }

\newcommand{\rf}[1]{(\ref{#1})}
\renewcommand{\theequation}{\thesection.\arabic{equation}}
\renewcommand{\thefootnote}{\fnsymbol{footnote}}
\newcommand{\newsection}{    
\setcounter{equation}{0}\section}
\def\appendix#1{\addtocounter{section}{1}\setcounter{equation}{0}
\renewcommand{\thesection}{\Alph{section}}
\section*{Appendix \thesection\protect\indent \parbox[t]{11.15cm}{#1}}
\addcontentsline{toc}{section}{Appendix \thesection\ \ \ #1}}

\def\nline{\,\nabla\kern -0.7em\raise0.2ex\hbox{/}\,\,}
\def\yline{\,y\kern -0.47em /}
\def\aline{\,a\kern -0.49em /}
\def\parline{\,\partial\kern -0.55em /\,\,}

\def\be{\begin{equation}}
\def\ee{\end{equation}}
\def\beq{\begin{eqnarray}}
\def\eeq{\end{eqnarray}}

\def\sm3{{\scriptscriptstyle (3)}}

\def\a{\alpha}

\def\vp{\vphantom{5pt}}

\def\as{{\a\kern-0.35em\raise0.95ex\hbox{${\scriptscriptstyle *}$}}}

\def\hbf{{\bf h}}

\def\Ys{{Y\kern-0.5em\raise1.5ex\hbox{${\scriptscriptstyle *}$}}}
\def\hs{{{\sf h}\kern -0.45em\raise 2ex
\hbox{$\scriptscriptstyle*$}}\,{}_{>0}^{\vp}}

\def\HG#1#2{{#1\kern -0.55em\raise 2.1ex\hbox{$\scriptscriptstyle (#2)$}}}

\def\OO{{\cal O}}

\begin{document}


\begin{flushright}
FIAN/TD/11-05 \\
hep-th/0512330
\end{flushright}

\vspace{1cm}

\begin{center}

{\Large \bf Light-cone formulation of  conformal  field theory

\bigskip adapted to AdS/CFT correspondence}

\vspace{2.5cm}

R.R. Metsaev\footnote{ E-mail: metsaev@lpi.ru }

\vspace{1cm}

{\it Department of Theoretical Physics, P.N. Lebedev Physical
Institute, \\ Leninsky prospect 53,  Moscow 119991, Russia }

\vspace{3.5cm}

{\bf Abstract}

\end{center}

Light-cone formulation of conformal field theory in space-time of
arbitrary dimension is developed. Conformal fundamental and shadow
fields with arbitrary conformal dimension and arbitrary spin are
studied. Representation of conformal algebra generators on space of
conformal fundamental and shadow fields in terms of spin operators
which enter in light-cone gauge formulation of field dynamics in
$AdS$ space is found. As an example of application of light-cone
formalism we discuss $AdS/CFT$ correspondence for massive arbitrary
spin $AdS$ fields and corresponding boundary $CFT$ fields at the
level of two point function.

\newpage
\renewcommand{\thefootnote}{\arabic{footnote}}
\setcounter{footnote}{0}

\section{Introduction}

Conjectured duality \cite{Maldacena:1997re} of large $N$ conformal
${\cal N}=4$ SYM theory and type IIB superstring theory in $AdS_5
\times S^5$ has triggered intensive study of field (string) dynamics
in AdS space and conformal field theory. By now it is clear that in
order to understand the conjectured duality better it is necessary to
develop powerful approaches to study of field (string) dynamics in
AdS space as well as conformal field theory. Light-cone approach is
one of the promising approaches which might be helpful to understand
AdS/CFT duality better. As is well known, quantization of
Green-Schwarz superstrings propagating in flat space is
straightforward only in the light-cone gauge. Since, by analogy with
flat space, we expect that a quantization of the Green-Schwarz $AdS$
superstring with a Ramond - Ramond charge \cite{Metsaev:1998it} will
be straightforward only in the light-cone gauge \cite{mt3}, it seems
that from the stringy perspective of $AdS/CFT$ correspondence the
light-cone approach to conformal field theory is the fruitful
direction to go. Light-cone formulation of {\it totally symmetric
conformal fundamental fields with canonical conformal dimension} and
{\it  shadow fields} was obtained in \cite{Metsaev:1999ui}. In this
letter we develop light-cone formulation for {\it conformal
fundamental and shadow fields with arbitrary conformal dimension and
arbitrary type of symmetry} (totally symmetric and mixed symmetry).

Let us first formulate the main problem we solve in this letter.
Conformal fundamental fields in $d-1$ dimensional space-time are
associated with representations of $SO(d-1,2)$ group labelled by
$\Delta$, eigenvalue of the dilatation operator, and by ${\bf
h}=(h_1,\ldots h_\nu)$, $\nu=[\frac{d-1}{2}]$, which is the highest
weight of the unitary representation of the $SO(d-1)$ group. The
highest weights $h_i$ are integers and half-integers for bosonic and
fermionic fileds respectively. The conformal dimension $\Delta$ and
${\bf h}$ satisfy the restriction
\be\label{uniresrt} \Delta \geq \Delta_0\,,\ee
where the canonical conformal dimension $\Delta_0$ is given by%
\footnote{Unitarity restriction \rf{unitrest} of the $so(d-1,2)$
algebra for the cases $d=4,5$ was studied in \cite{Evans,Mack:1975je}
and for arbitrary $d$ in \cite{Metsaev:1995re}. For discussion of
unitary representations of various superalgebras see
e.g.\cite{Dobrev:1985qv,Gunaydin:1998sw}.}
\be \label{unitrest}  \Delta_0 = h_k  - k  - 2 + d\,, \ee
and a number $k$ is defined from the relation%
\footnote{The labels $h_i$ are the standard Gelfand-Zeitlin labels.
They are related with Dynkin labels $h_i^D$ by formula:
$(h^D_1,h^D_2,\ldots,h^D_{\nu-1},h^D_\nu)=
(h_1-h_2,h_2-h_3,\ldots,h_{\nu-1}-h_\nu, h_{\nu-1}+h_\nu)$.}
\be\label{mineq2} h^{}_1=\ldots = h_k^{} > h_{k+1}^{} \ge
h^{}_{k+2}\ge \ldots \ge h_\nu^{}\geq 0, \ee
where for odd $d$ the weight $h_\nu$ should be replaced by $|h_\nu|$
(see e.g.\cite{Mack:1975je}). Another representations of the
conformal group associated with so called shadow fields have
eigenvalue of dilatation operator $\tilde\Delta$ given by
\be \tilde\Delta = d-1 - \Delta\,. \ee
Bosonic (fermionic) fields with ${\bf h}=(h_1,0,\ldots,0)$ (${\bf
h}=(h_1,1/2,\ldots,1/2,h_\nu)$, $h_\nu=\pm 1/2$ for odd $d$, $h_\nu
=1/2$ for even $d$) are referred to as totally symmetric conformal
fundamental and shadow fields
\footnote{We note that $\Delta= d-2$, ${\bf h}=(1,0,\ldots,0)$ and
$\Delta = d-1$, ${\bf h}=(2,0,\ldots,0)$ correspond to conserved
vector current and conserved traceless spin two tensor field
(energy-momentum tensor) respectively. Conserved conformal currents
can be built from massless scalar, spinor and spin 1 fields (see e.g.
\cite{Konstein:2000bi}). The shadow fields with $\tilde\Delta=2-s$,
${\bf h}=(s,0,\ldots,0)$ can be used to formulate higher derivatives
conformally invariant equations of motion for spin $s$ totally
symmetric tensor fields (see e.g.
\cite{Fradkin:1985am}-\cite{Segal:2002gd}). Discussion of conformally
invariant equations for mixed symmetry tensor fields with discrete
$\Delta$ may be found in \cite{Shaynkman:2004vu}.}.
In manifestly Lorentz covariant formulation the bosonic(fermionic)
totally symmetric conformal and shadow fields are described by a set
of the tensor (tensor-spinor) fields whose $SO(d-2,1)$ space-time
tensor indices have the structure of the respective Young tableauxes
with one row. Lorentz covariant description of totally symmetric
arbitrary spin conformal fundamental fields with $\Delta=\Delta_0$
and shadow fields with $\tilde\Delta=d-1-\Delta_0$ is well known.
Light-cone description of such fields was developed in
\cite{Metsaev:1999ui}. Bosonic (fermionic) fields with $h_2>0$
($|h_2|>1/2$) are referred to as mixed symmetry fields. In this paper
we develop light-cone formulation of CFT \cite{Metsaev:1999ui} which
is applicable to description of mixed symmetry conformal fundamental
(and shadow) fields with arbitrary conformal dimension $\Delta >
\Delta_0$ and for arbitrary space dimensions.

Remarkable feature of light-cone approach to CFT we exploit is that
number of physical spin D.o.F. of massless field in $d$-dimensional
space-time coincides with the number of independent spin degrees of
freedom of the corresponding conformal fundamental field (and shadow
field) with $\Delta = \Delta_0$ in $d-1$ -dimensional space-time%
\footnote{For the case of totally symmetric tensor fields this simple
fact can be checked in a rather straightforward way (see
e.g.\cite{Metsaev:1999ui}). It is naturally to expect that light-cone
matching of spin degrees of freedom is still to be case for the case
of mixed symmetry AdS massless fields (which are beyond scope of this
paper) and corresponding conformal fundamental and shadow fields.}.
The same coincidence holds true for massive field in $d$-dimensional
space and the corresponding conformal fundamental field (and shadow
field) in $d-1$-dimensional space with anomalous conformal dimension,
$\Delta
> \Delta_0$. It is this fact that allows us to develop the
representation for generators of the conformal algebra $so(d-1,2)$
acting in space of conformal fundamental and shadow fields in terms
of spin operators which enters in light-cone gauge description of
fields propagating in AdS space time. This is a reason why we refer
to such representation for CFT generators as AdS friendly
representation.

\newsection{Light-cone from of field dynamics in AdS space}

In order to demonstrate explicit parallel between light-cone
formulation of field dynamics in AdS space and that of conformal
field theory we begin with discussion of light-cone formulation of
field dynamics in AdS space developed in
\cite{Metsaev:1999ui,{Metsaev:2003cu}}. Let $\phi(x)$ be a bosonic
arbitrary spin field propagating in $AdS_d$ space. If we collect spin
degrees of freedom in a ket-vector $|\phi\rangle$ then a light-cone
gauge action for $\phi$ can be cast into the following form\cite{Metsaev:1999ui}%
\footnote{ We use parametrization of $AdS_d$ space in which
$ds^2=(-dx_0^2+dx_i^2+dx_{d-1}^2+dz^2)/z^2$. Light-cone coordinates
in $\pm$ directions are defined as $x^\pm=(x^{d-1} \pm x^0)/\sqrt{2}$
and $x^+$ is taken to be a light-cone time. We adopt the conventions:
$\partial^i=\partial_i\equiv\partial/\partial x^i$,
$\partial_z\equiv\partial/\partial z$, $\partial^\pm=\partial_\mp
\equiv \partial/\partial x^\mp$, $z\equiv x^{d-2}$ and use indices
$i,j =1,\ldots, d-3$; $I,J=1,\ldots, d-2$. Vectors of $so(d-2)$
algebra are decomposed as $X^I=(X^i,X^z)$.}
\be\label{lcact} S_{l.c.} =\frac{1}{2}\int d^dx \langle
\phi|\bigl(\Box -\frac{1}{z^2}A\bigr)|\phi\rangle\,, \qquad \Box =
2\partial^+\partial^- + \partial^i\partial^i + \partial_z^2\,. \ee
An operator $A$ being independent of space-time coordinates and their
derivatives is referred to as $AdS$ mass operator. This operator acts
only on spin indices of $|\phi\rangle$.

We turn now to discussion of global $so(d-1,2)$ symmetries of the
light-cone gauge action. The choice of the light-cone gauge spoils
the manifest global symmetries,  and in order to demonstrate that
these global invariances are still present one needs to find the
Noether charges which generate them. Noether charges (or generators)
can be split into kinematical and dynamical generators.  In this
paper we deal with free fields. At a quadratic level both kinematical
and dynamical generators have the following standard representation
in terms of the physical light-cone field
\be\label{hatG} \hat{G}=\int
dx^-d^{d-2}x\langle\partial^+\phi|G|\phi\rangle\,. \ee
Representation for the kinematical generators in terms  of
differential operators $G$ acting on the physical field
$|\phi\rangle$ is given by
\beq
\label{pi}&& P^i=\partial^i\,, \qquad P^+=\partial^+\,,
\\
&& D=x^+ P^- +x^-\partial^++x^I\partial^I+\frac{d-2}{2}\,,
\\
&& J^{+-}=x^+ P^- -x^-\partial^+\,,\qquad
\\
&& J^{+i}=x^+\partial^i-x^i\partial^+\,,
\\
&& J^{ij} = x^i\partial^j-x^j\partial^i + M^{ij}\,,
\\
\label{Kpads}&& K^+ = -\frac{1}{2}(2x^+x^-+x^Jx^J)\partial^+ +
x^+D\,,
\\
&& K^i = -\frac{1}{2}(2x^+x^-+x^Jx^J)\partial^i +x^i
D+M^{iJ}x^J+M^{i-}x^+\,, \eeq
while a representation for the dynamical generators takes the form
\beq && P^-=-\frac{\partial^I\partial^I}{2\partial^+}
+\frac{1}{2z^2\partial^+}A\,,
\\
&& J^{-i}=x^-\partial^i-x^i P^-
+M^{-i}\,,\\
\label{km}&& K^-=-\frac{1}{2}(2x^+x^-+x_I^2) P^- +
x^-D+\frac{1}{\partial^+}x^I\partial^JM^{IJ}
-\frac{x^i}{2z\partial^+}[M^{zi},A] +\frac{1}{\partial^+}B\,,\ \ \ \
\ \eeq
where $M^{-i}=-M^{i-}$ and
\be M^{-i} \equiv M^{iJ}\frac{\partial^J}{\partial^+}
-\frac{1}{2z\partial^+}[M^{zi},A]\,. \ee
Operators $A$, $B$, $M^{IJ}$ are acting only on spin degrees of
freedom of the field $|\phi\rangle$. $M^{IJ}=M^{ij},M^{zi}$ are spin
operators of the $so(d-2)$ algebra
\be\label{d2comrel} [M^{IJ},M^{KL}]=\delta^{JK}M^{IL} +3 \hbox{
terms}\,,\qquad M^{IJ}{}^\dagger = - M^{IJ}\,,\ee
while the operators $A$ and $B$ admit the following representation
\beq \label{adsope} A & = & 2B^z+ 2 M^{zi}M^{zi} +
\frac{1}{2}M^{ij}M^{ij} + \langle C_{AdS}\rangle +\frac{d(d-2)}{4}
\,,
\\
\label{bope} B & = & B^z + M^{zi}M^{zi}\,. \eeq
$\langle C_{AdS}\rangle$ is eigenvalue of the second order Casimir
operator of the $so(d-1,2)$ algebra for the representation labelled
by $D(E_0,{\bf h})$:
\be\label{casope1} \langle C_{AdS}\rangle =E_0(E_0+1-d)+
\sum_{\sigma=1}^\nu h_\sigma (h_\sigma -2\sigma +d-1)\,,\ee
while $B^z$ is $z$-component of $so(d-2)$ algebra vector $B^I$ which
satisfies the  defining equation%
\footnote{We use the notation $(M^3)^{[I|J]}\equiv
\frac{1}{2}M^{IK}M^{KL}M^{LJ} -(I\leftrightarrow J)$, $M^2\equiv
M^{IJ}M^{IJ}$.}
\be\label{basequ0} [B^I,B^J] =  \Bigl(\langle C_{AdS}\rangle +
\frac{1}{2}M^2 + \frac{d^2- 5d + 8}{2}\Bigr)M^{IJ} - (M^3)^{[I|J]}
\,. \ee
Equation \rf{basequ0} is a basic equation of light-cone gauge form of
the relativistic dynamics in AdS space. General method of solving
this equation may be found in Ref.\cite{Metsaev:2004ee}. As was noted
the operator $B^I$ transforms in vector representation of the
$so(d-2)$ algebra
\be\label{bId2tra} [B^I,M^{JK}]=\delta^{IJ}B^K-\delta^{IK}B^J\,.\ee
One can check then that the light-cone gauge action \rf{lcact} is
invariant with respect to the global symmetries generated by
$so(d-1,2)$ algebra taken to be in the form $\delta_{\hat{G}}
|\phi\rangle = G|\phi\rangle$.

As seen AdS mass operator plays important role in light-cone
approach. This operator is fixed by equations
\rf{adsope},\rf{basequ0}. In study of AdS/CFT correspondence it is
desirable to know explicit diagonalized operator $A$ taken to be in
the form
\be\label{Akappa} A = \kappa^2 - \frac{1}{4} \,.\ee
The explicit diagonalized form of the operator $\kappa$ is known for
the following cases:\\
i) arbitrary spin totally symmetric and antisymmetric massless fields
in $AdS_d$ \cite{Metsaev:1999ui};\\
ii) type IIB supergravity in $AdS_5 \times S^5$ and $AdS_3 \times
S^3$ backgrounds \cite{Metsaev:1999gz,Metsaev:2000mv};\\
iii) mixed symmetry arbitrary spin massless and self-dual massive
fields in $AdS_5$ \cite{Metsaev:2002vr};

We discuss now operator $\kappa$ for mixed symmetry arbitrary spin
massive field in $AdS_d$. Massive field $|\phi\rangle$ associated
with representation $D(E_0,\hbf)$ transforms in irreps of $so(d-1)$
algebra where $\hbf$  is the highest weight of the unitary
representation of the $so(d-1)$ algebra. This field can be decomposed
into representation of $so(d-3)\times so(2)$ subalgebra as follows
(see formula (5.38) in \cite{Metsaev:1999ui})
\be\label{phisod3dec} |\phi\rangle = \sum_{s'} \oplus
|\phi_{s'}\rangle \qquad
s' = \left\{\begin{array}{ll} 0\,,\pm 1, \ldots ,\pm h_k\,, &
\hbox{for bosonic fields},
\\[5pt]
\pm \frac{1}{2}, \ldots , \pm h_k\,,& \hbox{for fermionic fields},
\end{array}\right.
\ee
where $|\phi_{s'}\rangle$ are representations%
\footnote{For the case of totally symmetric massive bosonic
(fermionic) fields the $|\phi_{s'}\rangle$ is irreducible spin $|s'|$
($|s'|-\frac{1}{2}$) tensor (tensor-spinor) of the $so(d-3)$ algebra.
For mixed symmetry fields the $|\phi_{s'}\rangle$ is reducible
representation of the $so(d-3)$ algebra in general.}
of the $so(d-3)$ algebra and $s'$ is eigenvalue of the $so(2)$
algebra generator. On the whole space of $|\phi_{s'}\rangle$ the
operator $\kappa$ takes eigenvalue
\be\label{kappaeig} \kappa_{s'} = E_0 - \frac{d-1}{2} + s'\,,\ee
where lowest energy value $E_0$ is expressible in terms of standard
mass parameter as \cite{Metsaev:2003cu}:
\beq \label{bose0m}&& E_0 = \frac{d-1}{2} + \sqrt{ m^2 + \Bigl(h_k -k
+\frac{d-3}{2}\Bigr)^2\,}\,,\hspace{1.5cm} \hbox{ for bosonic
fields}; \ \
\\[5pt]
\label{fere0m}&& E_0 = m + h_k -k -2 +d\,,\,\,\, \hspace{4.3cm}
\hbox{ for fermionic fields}\,,  \ \ \eeq
and the number $k$ is defined from the relation \rf{mineq2}.

\newsection{Light-cone form of conformal field theory}

We present now light-cone formulation of conformal field theory. In
Ref.\cite{Metsaev:1999ui} we have developed light-cone formulation of
totally symmetric conformal fundamental fields with canonical
conformal dimension $\Delta_0$ and associated shadow fields starting
with Lorentz covariant formulation. This strategy is difficult to
realize in many cases because the Lorentz covariant formulations are
not available in general. One of attractive features of light-cone
formalism is that it admits to formulate CFT without knowledge of
Lorentz covariant formulation. The practice we have got while
deriving light-cone formulation of totally symmetric conformal
fundamental and shadow fields allows us to develop general light-cone
formalism. In this section we construct light-cone form of the
$so(d-1,2)$ conformal algebra generators for arbitrary conformal
dimension $\Delta$ \rf{uniresrt} and arbitrary symmetry (totally
symmetric and mixed symmetry) conformal fundamental and shadow
fields. We show that these generators can be constructed in terms of
the spin operators and the AdS mass operator which appear in
light-cone gauge formulation of field dynamics in AdS space. We start
with discussion of bosonic fields.

To develop general light-cone we should make an assumption about
generators. Based on our previous study of totally symmetric
conformal fundamental and shadow fields (see Appendix C of
Ref.\cite{Metsaev:1999ui}) we make the following two assumptions
about structure of the conformal
algebra generators acting on conformal fundamental and shadow fields%
\footnote{Commutation relations of the $so(d-1,2)$ algebra we use may
be found in Eqs.(2.3)-(2.6) of Ref.\cite{Metsaev:1999ui}.} .

{\bf (i)} Taking into account the form of the generators $P^a$,
$J^{ab}$, $D$ found for the case of totally symmetric representations
we suppose that they maintain this form for arbitrary representations
\beq \label{post01} && P^a = \partial^a\,,
\\
 \label{post02}  && J^{+i}=l^{+i}\,,
\\
 \label{post03}  && J^{+-}=l^{+-}\,,
\\
 \label{post04}  && J^{ij}=l^{ij} + M^{ij}\,,
\\
 \label{post05}  && J^{-i} = l^{-i} + M^{-i}\,,
\\
\label{post06}  && D = x^a\partial_a + \frac{d-2}{2}\,, \eeq
where we use the notation%
\footnote{Lorentz basis coordinates of $d-1$ dimensional flat space
$x^a$, $a=0,1,\ldots, d-3,d-1$ are decomposed in light-cone basis
coordinates $x^\pm$, $x^i$, $i=1,\ldots d-3$, $x^\pm \equiv (x^{d-1}
\pm x^0)/\sqrt{2}$. The derivatives $\partial_a\equiv \partial_{x^a}$
are decomposed as $\partial_+ = \partial_{x^+}=
\partial^-$, $\partial_- = \partial_{x^-}= \partial^+$ and
$\partial_i = \partial_{x^i}= \partial^i$.}
\beq
&& l^{ab} \equiv x^a \partial^b - x^b \partial^a\,,
\\
\label{K0mu} && K_0^a \equiv -\frac{1}{2}x^2 \partial^a + x^a x^b
\partial_b\,,
\\
&& M^{-i} \equiv  M^{ij} \frac{\partial^j}{\partial^+} +
\frac{q}{\partial^+} M^i\,,
\\
&& q \equiv \sqrt{\partial_a
\partial^a }\,,
 \qquad x^2 \equiv x_a x^a \,.\eeq
$M^{ij}$ and $M^i$ are the spin operators, i.e. they are acting only
on the spin degrees of freedom of CFT fields. The operators $M^{ij}$
satisfy commutators of the $so(d-3)$ algebra, while the operators
$M^{ij}$ and  $M^i$ satisfy commutators of the $so(d-2)$ algebra:
$$ [M^{ij}, M^{kl}] = \delta^{jk} M^{il} + 3 \hbox{ terms}\,, \qquad
[M^i, M^j]=M^{ij}\,, $$
\be [M^i, M^{jk}] = \delta^{ij} M^k - \delta^{ik}M^j\,. \ee
These operators subject the following hermitian conjugation rules
\be \label{herpromcft} M^{ij}{}^\dagger = - M^{ij}\,, \qquad
M^i{}^\dagger = M^i\,.\ee

{\bf (ii)} The generator $K^+$ has the following form
\be\label{Kp01}  K^+ = K_0^+ + \frac{d-2}{2}x^+
-\frac{\partial^+}{2q^2}A\,.\ee

Note that we do not make assumptions about the form of the remaining
generators $K^i$ and $K^-$. Now the problem we solve is formulated as
follows. Given the spin operators $M^{ij}$, $M^i$ find the operator
$A$ \rf{Kp01} and the remaining generators $K^i$, $K^-$. In Appendix
A by exploiting only the commutation relations of the conformal
algebra $so(d-1,2)$ we demonstrate that the above assumptions turn
out to be sufficient to find the remaining generators and to get
closed defining equations for the operator $A$. We present our final
result for the generators $K^i$ and $K^-$
\beq
\label{Kigen01}  K^i & = & K_0^i + \frac{d-2}{2}x^i
-\frac{\partial^i}{2q^2}A + M^{ij}x^j - x^+ M^{-i} + \frac{1}{2q}
[M^i,A]\,,
\\
\label{Kmgen01}  K^-  & = & K_0^- + \frac{d-2}{2}x^-
-\frac{\partial^-}{2q^2}A +\frac{1}{\partial^+}(M^{ij}x^i\partial^j +
M^i x^i q)
\nonumber\\
& - & \frac{\partial^i}{2q\partial^+}[M^i,A]
+\frac{1}{\partial^+}B\,, \eeq
where we introduce new spin operator $B$. By definition this operator
acts only on the spin degrees of freedom of conformal fundamental and
shadow fields. As in the case of the light-cone gauge formulation of
fields dynamics in AdS space the operators $A$ and $B$ admit the
representation
\beq \label{adsopecft} A & = &  2B^{0'} - 2 M^i M^i +
\frac{1}{2}M^{ij}M^{ij} + \langle C_{_{CFT}} \rangle
+\frac{d(d-2)}{4}\,,
\\
\label{bopecft} B & = & B^{0'} - M^i M^i\,, \eeq
where vector $B^I$ is decomposed as $B^I = (B^{0'}, B^i)$ and
satisfies the basic equation%
\footnote{In this section the `transverse' indices $I =\{0',i\}$, $J
=\{0',j\}$ are contracted by using flat metric $\eta^{IJ}$:
$\eta^{0'0'}=-1$, $\eta^{ij}=\delta^{ij}$. This is to say that
expression $M^2=M^{IJ}M^{IJ}$ takes the form $M^2= M^{ij}M^{ij} -
2M^iM^i$, i.e. we use the identification $M^{0'i}\equiv M^i$.}
\be\label{basequ0cft} - [B^I,B^J] = \Bigl(\langle C_{_{CFT}} \rangle
+ \frac{1}{2}M^2 + \frac{d^2 - 5d + 8}{2}\Bigr)M^{IJ}- (M^3)^{[I|J]}
\,. \ee
$\langle C_{CFT}\rangle$ is eigenvalue of the second order Casimir
operator of the $so(d-1,2)$ algebra for the representation labelled
by $\Delta$ and ${\bf h}$. The $\langle C_{CFT}\rangle$ is obtainable
from \rf{casope1} where we should make the replacement $E_0
\rightarrow \Delta$.

Comparing Eqs.\rf{adsope}-\rf{basequ0} and
\rf{adsopecft}-\rf{basequ0cft} we conclude that the operators $A$ and
$B$ in light-cone form of CFT satisfy the same equations as in
light-cone
gauge formulation of field dynamics in AdS space%
\footnote{Some differences in signs in Eqs.\rf{adsope}-\rf{basequ0}
and Eqs.\rf{adsopecft}-\rf{basequ0cft} are related with the fact we
use anti-hermitian spin operator $M^{zi}$ \rf{d2comrel} on AdS side
and hermitian $M^i$ \rf{herpromcft} on CFT  side.}.
We note also that on CFT and AdS sides these operators are realized
on fields having the same number of spin D.o.F. Because of the
representation for CFT generators given in \rf{post01}-\rf{Kmgen01}
is formulated in terms of the spin operators $A$, $B$, $M^{IJ}$ which
appear in light-cone gauge formulation of field dynamics in AdS space
we refer to this representation for CFT generators as AdS
friendly form of CFT%
\footnote{The relations \rf{pi}-\rf{km} and \rf{post01}-\rf{Kmgen01}
are applicable to the bosonic fields. Extension to the fermionic
fields is to make replacement $x^- \rightarrow x^-
+\frac{1}{2\partial^+}$ in these relations}.

Let us denote CFT field on which AdS friendly representation is
realized as $|\OO_{AdS}\rangle$. Generators acting on
$|\OO_{AdS}\rangle$ (see \rf{post05},\rf{Kp01}-\rf{Kmgen01}) are
non-polynomial with respect to the derivatives in transverse and
minus light-cone directions, $\partial^i$, $\partial^-$%
\footnote{Non-polynomial terms wit  respect to light-cone derivative
$\partial^+$ are unavoidable in light-cone formulation.}.
However, these non-polynomial terms can be cancelled by passing from
basis of $|\OO_{{AdS}}\rangle$ to the bases of conformal fundamental
fields and shadow fields which we shall denote by $|\OO\rangle$ and
$|\widetilde\OO\rangle$ respectively. We demonstrate cancellation of
non-polynomial terms in bases of $|\OO\rangle$ and
$|\widetilde\OO\rangle$ for the simplest case of generator $K^+$. To
this end we note that the AdS mass operator $A$ being hermitian can
be diagonalized on the whole space of $|\OO_{AdS}\rangle$. Being
diagonalized this operator can be presented in the form given in
\rf{Akappa}. Now we make the following transformation to the bases of
conformal fundamental and shadow fields
\be\label{OadsOO} |\OO_{AdS}\rangle = q^{-\omega}|\OO\rangle\,,
\qquad |\OO_{AdS}\rangle = q^{-\omega}|\widetilde{\OO}\rangle\,, \ee
where we use the notation

\be\label{sigdef} \omega = \left\{\begin{array}{cll}
\kappa + \frac{1}{2} & \hbox{ for conformal fundamental field} \ \ \
& |\OO\rangle\,,
\\[8pt]
- \kappa + \frac{1}{2} & \hbox{ for shadow field } \ &
|\widetilde\OO\rangle\,.
\end{array}\right. \ee

It easy to check that in the bases of conformal fundamental and
shadow fields $|\OO\rangle$, $|\widetilde\OO\rangle$ the generator
$K^+$ given in \rf{Kp01} takes the form

\be  K^+ = K_0^+ + (\omega +\frac{d-2}{2})x^+\,,\ee
where the respective values of $\omega$ are given in \rf{sigdef}.
Thus non-polynomial term $(1/q^2)A$ of expression $K^+$ \rf{Kp01}
cancels out in bases of $|\OO\rangle$ and $|\widetilde\OO\rangle$, as
it should be. We note also that in bases of $|\OO\rangle$ and
$|\widetilde\OO\rangle$ the dilatation operator is given by
\be D = x^a\partial_a + \omega +\frac{d-2}{2}\,, \ee
and the invariant scalar product on the space of $|\OO\rangle$ and
$|\widetilde\OO\rangle$  takes the standard form
\be (\widetilde\OO,\OO) = \int d^{d-1}x \langle
\widetilde\OO||\OO\rangle \,.\ee

\newsection{AdS/CFT correspondence}

After we have derived the light-cone formulation for  both the bulk
massive arbitrary spin AdS fields and the boundary CFT fields we are
ready to demonstrate explicitly AdS/CFT correspondence. Euclidean
version of this correspondence for various particular cases has been
studied in \cite{Aref'eva:1998nn}--\cite{Polishchuk:1999nh}.
Intertwining operator realization of AdS/CFT correspondence was
investigated in \cite{Dobrev:1998md}. For review and complete list of
references see \cite{D'Hoker:2002aw}.  In this Section we study
AdS/CFT correspondence for both the Lorentzian and Euclidean
signatures.

{\bf Lorentzian signature}. We begin with study of correspondence for
AdS space of Lorentzian signature. Discussion of this correspondence
for the scalar field may be found in \cite{Balasubramanian:1998sn}
and for totally symmetric arbitrary spin massless fields in
\cite{Metsaev:1999ui}.

We demonstrate that boundary value of normalizable solution of bulk
equations of motion is related to the conformal fundamental field
$|\OO\rangle$, while that of non-normalizable solution is related to
the shadow field $|\widetilde\OO\rangle$ (see
\cite{Balasubramanian:1998sn,Metsaev:1999ui}). To this end we
consider light-cone equations of motion which take the form

\be
\Bigl(-\partial_z^2+\frac{1}{z^2}(\kappa^2-\frac{1}{4})\Bigr)|\phi\rangle
= q^2 |\phi\rangle\,, \ee
and obtain the following normalizable and non-normalizable solutions
\beq
&& \label{solequmot01} |\phi_{norm}(x,z)\rangle = Z_{\kappa}(qz) |
\OO_{AdS} (x)\rangle\,,
\\
\label{solequmot02} && |\phi_{non-norm}(x,z)\rangle = Z_{-\kappa}(qz)
|\OO_{AdS}(x)\rangle\,, \eeq
where we use the notation $Z_\kappa(z)\equiv \sqrt{z}J_\kappa(z)$
and $J_\kappa$ is Bessel function%
\footnote{To keep discussion from becoming unwieldy here we restrict
our attention to non-integer $\kappa$. In this case the solutions
given in \rf{solequmot01},\rf{solequmot02} are independent.}.
In \rf{solequmot01} we use the notation for the boundary field
$|\OO_{AdS}\rangle$ since we are going to demonstrate that this field
is indeed a carrier of AdS friendly  representation discussed above.
This is to say that AdS transformations for bulk field $|\phi\rangle$
lead to conformal theory transformations for boundary field
$|\OO_{AdS}\rangle$

\be \label{gadsgcft} G_{_{AdS}}|\phi_{norm}(x,z)\rangle  =
Z_\kappa(qz) G_{_{CFT}} | \OO_{AdS}(x)\rangle\,. \ee
Here and below we use the notation $G_{_{AdS}}$ and $G_{_{CFT}}$ to
indicate the realization of the $so(d-1,2)$ algebra generators on the
bulk field $|\phi\rangle$ \rf{pi}-\rf{km} and conformal theory field
$|\OO_{AdS}\rangle$ \rf{post01}-\rf{Kmgen01} respectively. Let us
demonstrate the matching \rf{gadsgcft} for the generator $K^+$. To
this end we use the following general relation
\beq (K_0^a -\frac{z^2}{2}\partial^a + x^a(y+z\partial_z))
Z_\kappa(qz) &=&Z_\kappa(qz)\Bigl(K_0^a
-\frac{\partial^a}{2q^2}(\kappa^2-\frac{1}{4}) +x^a (y + z
\partial_z )\Bigr)
\nonumber\\
\label{genreladscft} &+& \frac{\partial^a}{q}(\partial_q
Z_\kappa(qz)) (\frac{d-2}{2}- y - z\partial_z) \eeq
where $K_0^a$ is given in \rf{K0mu}. Adopting relation
\rf{genreladscft} for $y=\frac{d-2}{2}$, $a =+$, and using
expressions for $K_{_{AdS}}^+$ \rf{Kpads}, $K_{_{CFT}}^+$ \rf{Kp01},
and taking into account that $|\OO_{AdS}\rangle$ in \rf{solequmot01}
does not depend $z$ we get immediately that $K_{_{AdS}}^+$ and
$K_{_{CFT}}^+$ satisfy the relation \rf{gadsgcft}%
\footnote{For the case of the generators $P^a$, $J^{+i}$, $J^{+-}$,
$J^{ij}$, $D$ matching  \rf{gadsgcft} is obvious. This matching for
the generators $J^{-i}$, $K^i$, $K^-$ can be proved following
procedure explored in Section 7 of Ref.\cite{Metsaev:1999ui}.}.

Taking into account relations \rf{OadsOO} and
\rf{solequmot01},\rf{solequmot02} we make sure that for small $z$ one
has the local interrelations
\be \label{phiope} |\phi_{norm}(x,z)\rangle \sim
 z^{\kappa+\frac{1}{2}}|\OO(x)\rangle\,, \qquad |\phi_{non-norm}(x,z)\rangle \sim
z^{-\kappa+\frac{1}{2}} |\widetilde\OO(x)\rangle\,. \ee
i.e. it is the conformal fundamental field $|\OO\rangle$ (and shadow
field $|\widetilde\OO\rangle$) that is the boundary value of
$|\phi_{norm}\rangle$ (and $|\phi_{non-norm}\rangle$) as it should
be. Relations \rf{gadsgcft},\rf{phiope} explain desired AdS/CFT
correspondence in Lorentzian signature.

{\bf Euclidean signature}. Now we demonstrate AdS/CFT
correspondence in the Euclidean signature at the level of two point function%
\footnote{Discussion of $AdS/CFT$ correspondence for spin one Maxwell
field, $s=1$, and graviton, $s=2$, may be found in
\cite{Witten:1998qj} and \cite{Liu:1998bu,Arutyunov:1998ve}
respectively. Euclidean version of the AdS/CFT correspondence for
arbitrary spin massless fields was studied in \cite{Metsaev:2002vr}
(see also \cite{Germani:2004jf}).}.
In light-cone gauge the arbitrary spin massive and massless fields in
$AdS_d$ are described by respective  $so(d-1)$ and $so(d-2)$ tensor
fields. The Euclidean light-cone gauge action%
\footnote{Note that only in the remainder of this paper  we use the
Euclidean signature. For $d=4$ the $AdS$ mass operator is equal to
zero for all massless fields. Here we restrict our attention to the
dimensions $d\geq 5$.}
takes then the form
\be\label{acteuc} S_{l.c.}^E =  \frac{1}{2}\int d^dx\Bigl( \langle d
\phi| |d \phi\rangle  +\frac{1}{z^2} \langle  \phi| A
|\phi\rangle\Bigr)\,. \ee
Attractive feature of this action is that there are no contractions
of tensor indices of fields with those of space derivatives, i.e.,
the action looks like a sum of actions for `scalar' fields with
different mass terms. This allows us to extend the analysis of
Ref.\cite{Witten:1998qj} in a rather straightforward way. Using
Green's function method and the $AdS$ mass operator given in
\rf{Akappa}  a solution to equations of motion

\be \Bigl(-\partial_{\bf x}^2 -\partial_z^2
+\frac{1}{z^2}(\kappa^2-\frac{1}{4})\Bigr) |\phi({\bf x},z)\rangle =
0\,,\qquad {\bf x}\equiv (x^1,\ldots, x^{d-1})\ee
is found to be
\be |\phi({\bf x},z)\rangle  = \int d{\bf x}' \frac{z^{\kappa +
\frac{1}{2}}}{(z^2 +|{\bf x} -{\bf x}'|^2)^{\kappa + \frac{d-1}{2}}}
|\widetilde\OO({\bf x}')\rangle\,.\ee
As was expected (see \rf{solequmot02},\rf{phiope}) this solution
behaves for $z \rightarrow 0$ like $z^{-\kappa +
\frac{1}{2}}|\tilde\OO({\bf x})\rangle$. Plugging this solution into
the action \rf{acteuc} and evaluating a surface integral gives

\be\label{S-lc} S_{l.c.}^E = \frac{1}{2}\int d {\bf x}d {\bf x}'\
\langle\tilde{{\cal O}}({\bf x})| \frac{\kappa + \frac{1}{2}}{|{\bf
x} -{\bf x}'|^{2\kappa + d-1}}|\tilde{{\cal O}}({\bf
x}')\rangle\,.\ee
This is light-cone representation for  two point function of
conserved current. The $so(d-3)$ decomposition of \rf{S-lc} can be
obtained by using \rf{phisod3dec},\rf{kappaeig}.

{\bf Conclusions}. The results presented here should have a number of
interesting applications and generalizations, some of which are:

i) In this paper we develop light-cone formulation for conformal
fundamental fields and shadow fields. As is well known shadow fields
can be used to formulate conformal invariant equations of motion. It
would be interesting to apply light-cone approach to study of
conformal invariant equations of motion for mixed symmetry conformal
fields.

ii) The light-cone formulation we develop in this paper allows us to
study of bosonic and fermionic conformal fundamental (and shadow)
fields on an equal footing. It would be interesting to apply our
approach to study of supersymmetric multiplets of superconformal
algebra $psu(2,2|4)$ which are relevant in the study of AdS/CFT
correspondence.

\bigskip
{\bf Acknowledgments}. We thank O. Shaynkman and V. Zaykin for useful
comments. This work was supported by the INTAS project 03-51-6346, by
the RFBR Grant No.05-02-17217, RFBR Grant for Leading Scientific
Schools, Grant No. 1578-2003-2 and Russian Science Support
Foundation.

\setcounter{section}{0} \setcounter{subsection}{0}
\appendix{Derivation of generators $K^i$, $K^-$.}

In this appendix we outline procedure of derivation of representation
for generators $K^i$, $K^-$ given in \rf{Kigen01},\rf{Kmgen01} and
prove that the operators $A$, $B$ and the spin operators $M^{ij}$,
$M^i$ should satisfy the following equations
\beq \label{con01}&& [A,M^{ij}]=0 \,,
\\
&& \label{con1} 2\{M^i,A\}-[[M^i,A],A]=0\,,
\\
&& \label{con2} [M^i,[M^j,A]]-\{M^{il},M^{lj}\} - \{M^i, M^j\} = 2
\delta^{ij} B\,. \eeq
We proceed in the following way.

{\bf i}) We note that the commutation relations of $K^+$ with the
generators given in \rf{post01}-\rf{post06} imply that the operator
$A$ is independent of space-time coordinates $x^a$ and their
derivatives $\partial^a$ and commutes with spin operator $M^{ij}$,
i.e. we obtain \rf{con01}.

{\bf ii}) From commutator
\be [K^+, J^{-i}] = K^i  \ee
we find representation for $K^i$ given in \rf{Kigen01}.

{\bf iii}) Using expressions for $K^+$ \rf{Kp01} and $K^i$
\rf{Kigen01} we evaluate the commutator
\be  [K^+,K^i]= \frac{\partial^+}{4q^3}(-2\{M^i,A\}+[[M^i,A],A])\,.
\ee
From this relation and the commutator $[K^+,K^i]=0$ we obtain the
constraint \rf{con1}.

{\bf iv}) Making use of \rf{post05}  and \rf{Kigen01} we evaluate
then the commutator
\beq && [J^{-i},K^j]
\nonumber\\
&& =\delta^{ij} \Bigl(K_0^- + \frac{d-2}{2}x^-
-\frac{\partial^-}{2q^2}A +\frac{1}{\partial^+}(M^{kl} x^k \partial^l
+ M^l x^l q)-\frac{\partial^l}{2q\partial^+}[M^l,A]\Bigr)
\nonumber\\
&& + \frac{1}{2\partial^+} \Bigl([M^i, [M^j, A]] - \{M^{il}, M^{lj}\}
- \{M^i, M^j\} \Bigr)\,. \eeq
From this and the $so(d-1,2)$ algebra commutator $ [J^{-i},K^j] =
\delta^{ij}K^- $ we find the representation for the generator $K^-$
given in \rf{Kmgen01} provided the spin operators $M^{ij}$, the
operators $A$ and $B$ satisfy the constraint given in \rf{con2}. Note
that this constraint gives definition of operator $B$ in terms of
basic operators which are spin operator $M^{ij}$ and AdS mass
operator $A$.

Thus we obtain representation for generators $K^i$, $K^-$ given in
\rf{Kigen01},\rf{Kmgen01} and equations \rf{con01}-\rf{con2}. All
that remains is to prove that equations  \rf{con01}-\rf{con2} are
equivalent to representation for operators $A$ and $B$ and basic
defining equations given in \rf{adsopecft}-\rf{basequ0cft}. This can
be done by following a procedure explored in the Appendix A of
Ref.\cite{Metsaev:2003cu}.

\end{document}